\documentclass[a4paper,11pt]{article}
\pdfoutput=1 

\usepackage{jheppub} 

\usepackage[T1]{fontenc} 
\usepackage{physics}

\title{\boldmath Entanglement wedge cross-section in shock wave geometries}

\author{Jan Boruch}
\affiliation{Faculty of Physics, University of Warsaw, ul. Pasteura 5, 02-093 Warsaw, Poland}

\emailAdd{ja.boruch@student.uw.edu.pl}

\abstract{We consider reflected entropy in a thermofield double state perturbed by a heavy operator insertion. For sufficiently early operator insertions the dual geometry can be described by a localized shock wave geometry. We calculate the entanglement wedge cross-section in this geometry for symmetric intervals and find that it matches precisely with the CFT result for sufficiently late times. Our result exhibits a plateau before going to zero, a behaviour similar to the one observed recently in the context of global quantum quenches. We find that at high temperatures this behaviour is properly captured by the line-tension picture.}

\begin{document} 
\maketitle
\flushbottom

\section{Introduction}
\label{sec:intro}
The AdS/CFT correspondence~\cite{Maldacena_1999} greatly simplifies the study of the entanglement structure of quantum field theories. This is due to the geometrization of entanglement measures which manifests itself in the Ryu-Takayanagi formula~\cite{Ryu_2006,Hubeny_2007,Lewkowycz_2013,Dong_2016}.
In short, it states that for a CFT in a state $\ket{\Psi}$, the entanglement entropy of a subregion $A$
\begin{equation}
S_{A} = - \Tr \rho_{A} \log \rho_{A} , \phantom{aaa}
\rho_A = \Tr_{A^{c}} \ket{\Psi}\bra{\Psi} ,
\end{equation}
can be computed as the area of a codimension two surface in the bulk 
\begin{equation}
S_A = \frac{\textup{Area}(\gamma_A)}{4G} ,
\end{equation}
where $\gamma_A$ is the minimal surface anchored on $\partial A$ and homologous to region $A$.

For pure states entanglement entropy $S_A$, provides a good measure of quantum correlations, allowing us to probe the dynamics of entanglement under different types of quenches~\cite{Nozaki_2013,Caputa_2015,Calabrese_2005,Calabrese_2009,Caputa_2015_fin_temp,Asplund_2015_scrambling,Asplund_2015_heavystates}. For mixed states, however, von Neumann entropy measures both quantum and classical correlations, thus leading to a necessity of finding a different correlation measure, more suitable for mixed states.

Such a measure has been first proposed in~\cite{Umemoto_2018,Nguyen_2018} in the holographic context, with a precise holographic dual called entanglement wedge cross-section. Entanglement wedge cross-section is defined for two boundary intervals as an area of minimal cross-section which bipartitions their entanglement wedge 
\begin{equation}
E_W = \min \frac{\textup{Area}(\Sigma_{AB})}{4G} .
\end{equation}
This measure has been further related to other correlation measures~\cite{dutta2019canonical,Tamaoka_2019,Kudler_Flam_2019}. For recent developments see~\cite{Umemoto_2018_Multipartite,Umemoto_2019,akers2020entanglement,asrat2020tbart,Bao_2019_bits,Bao_2019_conditional,Bao_2019_multip_reflected,Bao_2019_wormholes,Harper_2019,Liu_2019,Jokela_2019,Velni_2019,Chu_2020,Du_2019,du2019inequalities,Ag_n_2019}. Let us now review the definitions of two measures dual to EWCS: entanglement of purification and reflected entropy.

\paragraph{Entanglement of purification.} Entanglement of purification~\cite{Terhal_2002} is defined as 
\begin{equation}
E_P (\rho_{AB}) = \min_{\rho_{AB}= \Tr_{A'B'}\ket{\psi}\bra{\psi}} S(\rho_{A A'}) ,
\end{equation}
where $\rho_{AA'}=\Tr_{BB'} \ket{\psi} \bra{\psi}$. The minimization is taken over all possible purifications of state $\rho_{AB}$, that is over all pure states $\ket{\psi} \in \mathcal{H}_{AA'} \otimes \mathcal{H}_{BB'}$ which satisfy the condition $\rho_{AB}= \Tr_{A'B'}\ket{\psi}\bra{\psi}$. The holographic entanglement of purification conjecture relates this quantity to entanglement wedge cross-section as~\cite{Umemoto_2018,Nguyen_2018}
\begin{equation}
E_W = E_P (\rho_{AB}) .
\end{equation}
Interesting realizations of this idea have been provided in~\cite{Hirai_2018,Caputa_2019}. In particular, in~\cite{Caputa_2019} the authors derived the holographic entanglement of purification as a simple entanglement entropy of a state deformed via path integral optimization procedure~\cite{Caputa_2017,Caputa_2017_PRL,Takayanagi_2018}. In general, however, this quantity is hard to compute on the field theory side. This is due to minimization over all possible purifications, which is not well understood in field theories.

\paragraph{Reflected entropy.} On the other hand, reflected entropy~\cite{dutta2019canonical} is defined as entanglement entropy of a canonically purified state. Consider a general mixed state 
\begin{equation}
\rho_{AB} = \sum_{a} p_a \ket{\psi_a}\bra{\psi_a} ,
\end{equation}
with $\sum_a p_a =1$, belonging to Hilbert space $\mathcal{H}_A \otimes \mathcal{H}_B$. States $\ket{\psi_a}\in \mathcal{H}_A \otimes \mathcal{H}_B $ can be written as
\begin{equation}
\ket{\psi_a} = \sum_{i} \sqrt{\lambda_a ^{i}} \ket{i_a}_A \otimes \ket{i_a}_B ,
\end{equation}
with $\sum_{i} \lambda_a ^i = 1$. With that our density matrix takes the form
\begin{equation}
\rho_{AB} = \sum_{a,i,j} p_a \sqrt{\lambda_a ^{i} \lambda_a ^{j}}
 \ket{i_a}_A \otimes \ket{i_a}_B \otimes  \bra{j_a}_A \otimes \bra{j_a}_B .
\end{equation}
Interpreting the above density matrix as a vector $\ket{\sqrt{\rho_{AB}}}\in \mathcal{H}_A \otimes \mathcal{H}_B \otimes \mathcal{H}_A ^* \otimes \mathcal{H}_B ^*$ defines a canonical purification 
\begin{equation}
\ket{\sqrt{\rho_{AB}}} = \sum_{a,i,j} p_a \sqrt{\lambda_a ^{i} \lambda_a ^{j}}
 \ket{i_a}_A \otimes \ket{i_a}_B \otimes  \ket{j_a}_{A^*} \otimes \ket{j_a}_{B^*} ,
\end{equation}
which satisfies the appropriate condition $\rho_{AB} = \Tr_{A^* B^*}{\ket{\sqrt{\rho_{AB}}}\bra{\sqrt{\rho_{AB}}}}$. Using the above state, we define reflected entropy as
\begin{equation}
S_R (A:B) = S(\rho_{AA^*}), \phantom{aa} 
\rho_{AA^{*}} = \Tr_{B B^*}{\ket{\sqrt{\rho_{AB}}}\bra{\sqrt{\rho_{AB}}}} .
\end{equation}
Reflected entropy is related to entanglement wedge cross-section via
\begin{equation}
S_R (A:B) = 2 E_W  .
\end{equation}
In contrast with the entanglement of purification, it is well understood how to compute reflected entropy in conformal field theories. In particular, there is clear replica trick prescription\footnote{See \cite{Jeong_2019} for a very clear presentation.} which can be used to compute it. This has been used in recent works to investigate the dynamical properties of reflected entropy in holographic CFTs~\cite{Kusuki_2020}\footnote{Dynamics of EWCS purely on the gravity side have been also studied in~\cite{Yang_2019,velni2020evolution}.} as well as in more general CFTs~\cite{Kudler_Flam_2020,moosa2020time}.
For our purposes, however, it will be sufficient to know that reflected entropy~\cite{dutta2019canonical} (or holographic entanglement of purification~\cite{Caputa_2019}) between two intervals $[z_1 ,z_2]$, $[z_3, z_4]$ in a 2d CFT on a complex plane can be expressed as 
\begin{equation}
E_W = \frac{c}{12} \log \frac{1+\sqrt{z}}{1-\sqrt{z}} +
\frac{c}{12} \log \frac{1+\sqrt{\bar{z}}}{1-\sqrt{\bar{z}}},
\end{equation}
with $z$ being the standard cross-ratio $z=z_{12}z_{34}/z_{13}z_{24}$.

In this paper, we provide another study of the dynamics of reflected entropy in holographic CFTs. 
The setup we consider is motivated by the works~\cite{Shenker_2014,Roberts_2015,Roberts_2015_shocks}. In particular, in \cite{Shenker_2014} the authors considered mutual information between two matching intervals on both sides of the thermofield double state, perturbed by a spherical shock wave. It was found that for sufficiently early perturbations mutual information between the two intervals goes to zero. Since the entanglement wedge cross-section is a different measure of correlations between two intervals, it is interesting to check what is the behaviour of the correlations captured by the entanglement wedge cross-section in this simple setup. We will find that for spherical shock waves the entanglement wedge cross-section behaves similarly to other correlation measures, such as mutual information and the two-point function. For localized shocks, however, we find an interesting behaviour in which the entanglement wedge cross-section settles at some non-zero value sometime before mutual information goes to zero. This results in a plateau, similar to the one observed in~\cite{Kudler_Flam_2020}. We show that at high temperatures this behaviour can be precisely captured by the line-tension picture~\cite{Nahum_2017,jonay2018coarsegrained,Mezei_2018,von_Keyserlingk_2018,Zhou_2019,Kudler_Flam_2020_linetension_negativity,Wang_2019}, which was very recently extended to the local operator quenches in~\cite{kudlerflam2020entanglement}.

The structure of the paper is as follows. In Section \ref{sec:CFT} we derive the result in (1+1)d CFT using two conformal maps. Our computation is similar in spirit to the computation of the two-point function in \cite{Roberts_2015}. In Section \ref{sec:EWCS_localized shock wave} we calculate the entanglement wedge cross-section in a localized shock wave geometry~\cite{Roberts_2015_shocks}. A precise match with a late time CFT result is found. We discuss some aspects of our result. In Section \ref{sec:EWCS spherical shock wave} we use the same techniques to compute EWCS for the case of a spherical shock wave~\cite{Shenker_2014}. In Section \ref{sec:discussion} we summarize our results.


\section{Reflected entropy for perturbed TFD from CFT}
\label{sec:CFT}
\begin{figure}
    \centering
    \includegraphics[width=15cm]{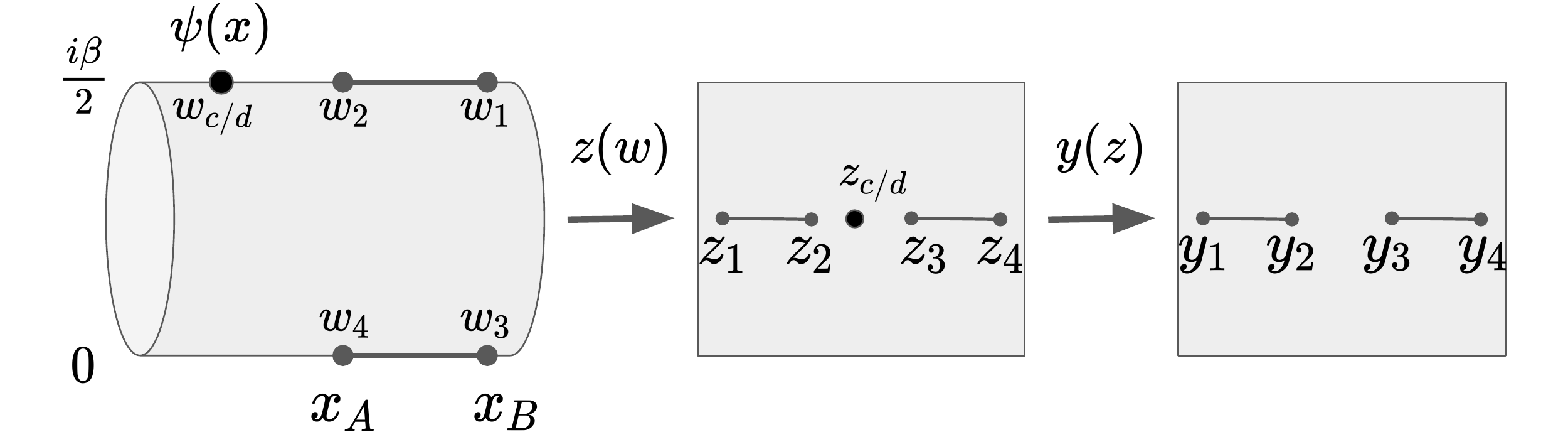}
    \caption{Our setup in the computation of reflected entropy. We consider two symmetric intervals $[x_A, x_B]$ in both copies of the CFT, perturbed by the heavy operator insertion in the left copy at time $t$ in the past. Using two conformal maps, $z(w)$ and $y(w)$, we map this setup to the known vacuum result.}
    \label{fig:CFT setup}
\end{figure}

We begin with the CFT side computation. The state of interest is a thermofield double state~\cite{Maldacena_2003,Hartman_2013,Morrison_2013} perturbed by a heavy operator insertion 
\begin{equation}
\ket{TFD}_{pert} = e^{-i H_L t} \psi_L (x) e^{i H_L t} 
\ket{TFD}. 
\end{equation}
One can think of it as first evolving the left CFT backward in time, inserting the operator $\psi(x)$, and then evolving the resulting state forward in time to $t_L=t_R=0$. We consider the case with symmetric intervals $[x_A , x_B]$ in both copies of CFT. To compute reflected entropy we need to find Euclidean path integral on a cylinder of circumference $\beta$, with twist operator insertions at points
\begin{align}
w_3 &= x_A, \phantom{a} \Bar{w}_3 = x_A,
\phantom{aaa,} w_4 = x_B,\phantom{a} \Bar{w}_4 = x_B,\\
w_1 &= x_B + \frac{i\beta}{2}, \phantom{aaa-x_B} 
\Bar{w}_1 = x_B - \frac{i\beta}{2}, \\ 
w_2 &= x_A + \frac{i\beta}{2}, \phantom{aaa-x_B}
\Bar{w}_2 = x_A - \frac{i\beta}{2} ,
\end{align}
and heavy operators inserted at
\begin{align}
w_c &= x-t-i \epsilon +  \frac{i \beta}{2}, \phantom{aa} 
\Bar{w}_c = x+t+i\epsilon -  \frac{i \beta}{2}, \\
w_d &= x-t + i\epsilon +  \frac{i \beta}{2}, \phantom{aa}
\Bar{w}_d = x+t-i\epsilon -  \frac{i \beta}{2},
\end{align}
where we introduced $\epsilon$ as the UV regulator of the local operator $\psi$. To do this most simply, we will first map our setup to the complex plane where the result is known~\cite{dutta2019canonical,Caputa_2019}
\begin{align}
E_W = \frac{S_R}{2} &= \frac{c}{6}\textup{arccosh} 
\left(
\frac{1+ \sqrt{u}}{\sqrt{\nu}}
\right)
, \\
u &= y \Bar{y}, \phantom{aa} \nu= (1-y)(1-\Bar{y}) ,
\label{eq:Ew-crossratios u,v}
\\
y&= \frac{(y_1 - y_2)(y_3-y_4)}{(y_1 - y_3)(y_2 - y_4)}, 
\end{align}
and then analytically continue it to Lorentzian times $-t$, such that $t>x_B-x$~\cite{Roberts_2015}.
We use two conformal maps (see Figure \ref{fig:CFT setup}).
The first map will take us from perturbed thermofield double state $(w,\Bar{w})$ to perturbed state on the complex plane $(z,\Bar{z})$~\cite{Calabrese_2009}
\begin{equation}
z(w) = e^{\frac{2\pi w}{\beta} } ,\phantom{aaa}
\Bar{z}(\Bar{w}) = e^{\frac{2\pi \Bar{w}}{\beta} },
\end{equation}
and the second one will take us further to the unperturbed complex plane $(y,\Bar{y})$~\cite{Asplund_2015_heavystates,Fitzpatrick_2014,Fitzpatrick_2015}
\begin{align}
y(z) &= \left( \frac{z-z_c}{z-z_d} \right)^{\alpha} ,
\phantom{aaa} \Bar{y}(\Bar{z}) = \left( \frac{\Bar{z}-\Bar{z}_c}{\Bar{z}-\Bar{z}_d} \right)^{\alpha} ,
\\
z_c & = -e^{\frac{2\pi}{\beta} (x-t-i \epsilon)}, 
\phantom{aaa} \Bar{z}_c =  -e^{\frac{2\pi}{\beta} (x+t+i \epsilon)}, \\
z_d & = -e^{\frac{2\pi}{\beta} (x-t+i \epsilon)}, 
\phantom{aaa} \Bar{z}_d =  -e^{\frac{2\pi}{\beta} (x+t-i \epsilon)} ,
\end{align}
where $\alpha = \sqrt{1 - \frac{24 h_\psi}{c}}\simeq 1-12 h_\psi/c$ for $h_\psi/c \ll 1$. Points $z_c$, $z_d$, correspond to the insertion points of $\psi, \psi^{\dagger}$ after using the first map. 
The "cross-ratios" necessary in (\ref{eq:Ew-crossratios u,v}) take the form 
\begin{equation}
y= \frac{\left( \left( \frac{z_1-z_c}{z_1-z_d} \right)^{\alpha}
- \left( \frac{z_2-z_c}{z_2-z_d} \right)^{\alpha}
\right)
\left( \left( \frac{z_3-z_c}{z_3-z_d} \right)^{\alpha}
- \left( \frac{z_4-z_c}{z_4-z_d} \right)^{\alpha}
\right)}
{\left( \left( \frac{z_1-z_c}{z_1-z_d} \right)^{\alpha}
- \left( \frac{z_3-z_c}{z_3-z_d} \right)^{\alpha}
\right)
\left( \left( \frac{z_2-z_c}{z_2-z_d} \right)^{\alpha}
- \left( \frac{z_4-z_c}{z_4-z_d} \right)^{\alpha} \right)},
\end{equation}
with analogic form for $\Bar{y}$. At this point, we perform an analytical continuation to Lorentzian times $t$. The terms appearing in the brackets above might have non-trivial monodromies around zero, which we need to properly take care of to get the correct answer. To analyze which of these terms pass through the branch cut during analytical continuation, we expand these terms for small $\epsilon$~\cite{Asplund_2015_heavystates,Caputa_2015}
\begin{align}
\frac{\bar{z}_1-\bar{z}_c}{\bar{z}_1-\bar{z}_d} &= 1+\frac{4 i \pi  \epsilon  e^{\frac{2 \pi  (t+x)}{\beta }}}{\beta  \left(e^{\frac{2 \pi  (t+x)}{\beta }}-e^{\frac{2 \pi  x_B}{\beta }}\right)}+O\left(\epsilon ^2\right) ,
\\
\frac{\bar{z}_2-\bar{z}_c}{\bar{z}_2-\bar{z}_d} &= 1+\frac{4 i \pi  \epsilon  e^{\frac{2 \pi  (t+x)}{\beta }}}{\beta  \left(e^{\frac{2 \pi  (t+x)}{\beta }}-e^{\frac{2 \pi  x_A}{\beta }}\right)}+O\left(\epsilon ^2\right)
.
\end{align}
We see that as we increase $t$ past $t=x_B -x$ and $t=x_A -x$, both of these expressions pass to a different sheet at infinity. The other terms stay on the principal sheet. We therefore take 
\begin{equation}
\frac{\bar{z}_1-\bar{z}_c}{\bar{z}_1-\bar{z}_d} \rightarrow \frac{\bar{z}_1-\bar{z}_c}{\bar{z}_1-\bar{z}_d} e^{2\pi i},\phantom{aaa} \frac{\bar{z}_2-\bar{z}_c}{\bar{z}_2-\bar{z}_d} \rightarrow
\frac{\bar{z}_2-\bar{z}_c}{\bar{z}_2-\bar{z}_d} e^{2\pi i} ,
\end{equation}
before taking $h_\psi/c \ll 1$. It is convenient to rewrite the full expression for $y$ as 
\begin{equation}
y= \frac{\left( 1
- \left( \frac{z_1-z_d}{z_1-z_c} \frac{z_2-z_c}{z_2-z_d} \right)^{1-12h_\psi/c}
\right)
\left( 1- \left( \frac{z_3-z_c}{z_3-z_d} \frac{z_4-z_d}{z_4-z_c} \right)^{1-12h_\psi/c}
\right)}
{\left( 1
- \left( e^{-2\pi i} \frac{z_1-z_d}{z_1-z_c} \frac{z_3-z_c}{z_3-z_d} \right)^{1-12h_\psi/c}
\right)
\left( 1- \left( e^{2\pi i} \frac{z_2-z_c}{z_2-z_d} \frac{z_4-z_d}{z_4-z_c} \right)^{1-12h_\psi/c}  \right)}.
\end{equation}
After expanding the terms in the brackets for $h_\psi/c \ll 1$ and assuming late times $t \gg x_B - x$, we arrive at
\begin{align}
\bar{y} &= \frac{\left( e^{\frac{2\pi}{\beta}x_A}-e^{\frac{2\pi}{\beta}x_B} \right)^2}
{ \left( e^{\frac{2\pi}{\beta}x_A} +e^{\frac{2\pi}{\beta}x_B}
+ \frac{24 h_\psi \pi i }{c \epsilon_\beta} e^{t+x}
\right)^2} , \phantom{aa} \epsilon_\beta \equiv (e^{- \frac{2\pi i\epsilon}{\beta}} - e^{\frac{2\pi i\epsilon}{\beta}}) ,
\\
y &= \tanh{\frac{\pi}{\beta}(x_A-x_B)}^2 .
\end{align}
This leads to 
\begin{align}
\label{eq:CFT result}
E_W &= \frac{c}{6} \textup{arccosh} \left(
\frac{2\cosh \frac{2\pi}{\beta}(x_A - x_B) + h(x_A) + h(x_B)}
{2 \sqrt{h(x_A)+1} \sqrt{h(x_B)+1}}
\right) ,\\
h(x_{A/B}) &= \frac{12h_\psi \pi i}{c \epsilon_{\beta}} 
e^{\frac{2\pi}{\beta}(t+x -x_{A/B})} .
\end{align}
In the next section, we will find that this result precisely matches the gravity calculation in localized shock wave geometry, after relating regularization parameters as $\epsilon=-\tau$. The important point of our CFT calculation was taking $t>x_B-x$. This agrees with the conclusion that local shock wave geometries are a good approximation in the region $t>\abs{x}$ derived in \cite{Roberts_2015_shocks}.

\section{Entanglement wedge cross-section for localized shock wave}
\label{sec:EWCS_localized shock wave}

\begin{figure}
    \centering
    \includegraphics[width=8cm]{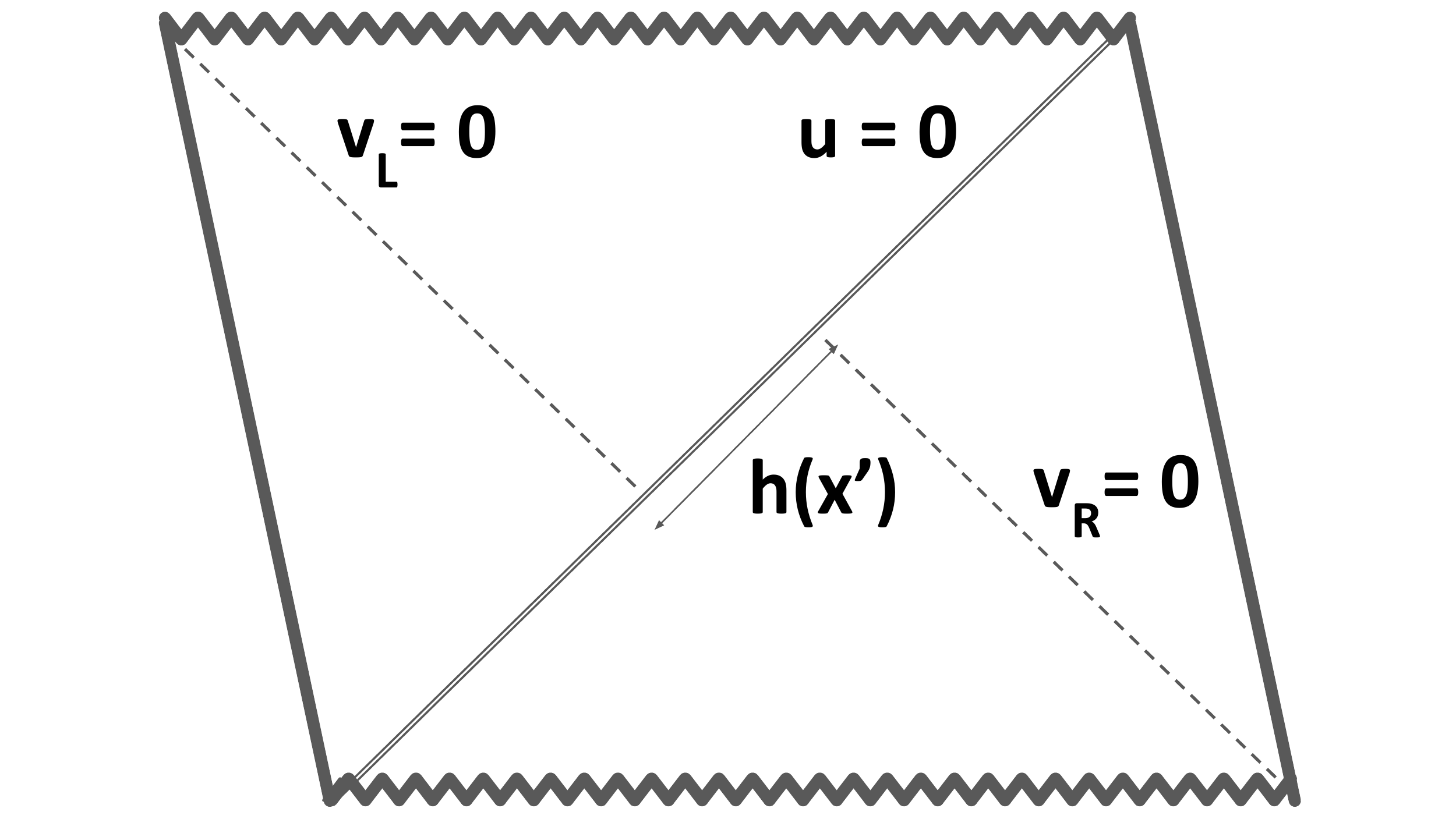}
    \caption{Two halves of eternal AdS black hole, glued together along the shock at $u=0$. There is a discontinuity in $v$ coodinate as we pass through the shock, with $v_L  = v_R  + h(x')$.}
    \label{fig:shockwave geometry}
\end{figure}

In this section, we proceed with the gravity side computation. For convenience we work with AdS radius $l=1$. We also set $R=1$ which will be reintroduced at the end of the computation. The holographic dual to thermofield double state perturbed by heavy operator insertion $h_\psi$ was found in~\cite{Roberts_2015,Roberts_2015_shocks}. It is a geometry of a localized shock wave
\begin{equation}
ds^2 = - \frac{4}{(1+uv)^2} du dv + 
\frac{(1-uv)^2}{(1+uv)^2} dx'^2 + 4 \delta(u) h(x') du^2 ,
\end{equation}
with 
\begin{equation}
h(x') = \frac{4 \pi G_N h_\psi }{\sin \tau}  e^{t-|x'-x|} . 
\end{equation}
In the above, the shock wave propagates from the point $(x,-t)$ on the left boundary and the parameter $\tau$ corresponds to the UV regularization of the single-particle operator in the bulk. This geometry can be understood as two halves of eternal AdS black hole, glued together along $u=0$ with a shift in $v$ coordinate (see Figure \ref{fig:shockwave geometry})
\begin{equation}
v_L  = v_R  + \delta v(x'),\phantom{aa} \delta v(x')= h(x') ,
\end{equation}
where we set coordinates $v_L$ to the future of the shock, and $v_R$ to the past. We will consider the case where $x<x_A<x_B$.

To find entanglement wedge cross-section between intervals $[x_A,x_B]$ on both boundaries, we first need to find the HRT surfaces between respective ends of the intervals. Following \cite{Roberts_2015}, we do this by calculating the distances from boundary points $(t_L=0, x_A)$ and $(t_R=0,x_A)$ to the intermediate point $(v_L=h(x_p)+v_R,x_p)$ on the horizon. Minimizing the total distance over the intermediate point will give us the HRT surface. The total distance can be found simply using embedding coordinates and reads\footnote{See Appendix \ref{app:conventions} for embedding coordinates.}
\begin{equation}
d  = \log \left(4 r_\infty ^2 \left(\cosh (x_A - x_p)- v_R \right) \left( (h(x_p)+v_R) +\cosh (x_A-x_p)\right)\right) .
\end{equation}
For the HRT surface anchored at the point $x_A$ the distance is minimized for 
\begin{align}
v_R (x_A) &= - \frac{h(x_A)}{2 \sqrt{h(x_A)+1}} , \\
x_p (x_A) &= x_A + \log \sqrt{h(x_A)+1} ,
\end{align}
with the total distance
\begin{equation}
d= \log \left( 4 r_\infty ^2 (1+h(x_A)) \right) .
\end{equation}
Knowing the intermediate points we can now calculate the entanglement wedge cross-section. Our approach will be similar to the one presented in \cite{Hirai_2018,Kusuki_2020}, with the slight generalization of the calculation to the case of bulk-boundary geodesics. Using embedding coordinates we will find the minimal distance between two bulk-boundary geodesics, each anchored on one interval endpoint and its respective intermediate point on the horizon.

\subsection{Spacelike geodesics in AdS}
A general spacelike geodesic anchored on bulk points $X_i$ and $X_j$ can be written as
\begin{equation}
X_{ij}^{A} (\lambda) = m^A e^{-\lambda} + n^A e^{\lambda} ,
\end{equation}
where $m^A, n^A$ satisfy 
\begin{equation}
m^A m_A = n^A n_A = 0, \phantom{aaa} 2 m \cdot n=-1 .
\label{eq:mn}
\end{equation}
Using 
\begin{equation}
X_{ij}^A (\lambda_i) = X_i ^{A}, 
\phantom{aaa} X_{ij}^A (\lambda_j) = X_j ^{A} ,
\end{equation}
we can express $X_{ij}(\lambda)$ as
\begin{equation}
X_{ij}^{A} (\lambda) = 
\frac{X_i ^A \sinh{(\lambda- \lambda_j)} - X_j ^A \sinh{(\lambda-\lambda_i)}}
{\sinh{(\lambda_i-\lambda_j)}}.
\end{equation}
Now we can use the equations (\ref{eq:mn}) to find 
\begin{align}
e^{\lambda_i} &= \sqrt{\xi - \sqrt{\xi^2 -1}}, \phantom{aaa}
e^{-\lambda_i} = \sqrt{\xi + \sqrt{\xi^2 -1}},
\\
e^{\lambda_j} &= \sqrt{\xi + \sqrt{\xi^2 -1}}, \phantom{aaa}
e^{-\lambda_j} = \sqrt{\xi - \sqrt{\xi^2 -1}}
,
\end{align}
where we introduced the product $\xi = -X_i \cdot X_j$. With that $X_{ij}(\lambda)$ can be rewritten further as
\begin{equation}
X_{ij}^A (\lambda) = C_- e^{-\lambda} + C_+ e^{\lambda} ,
\end{equation}
\begin{align}
C_- &= \frac{X_i ^A \sqrt{\xi + \sqrt{\xi^2 -1}} - X_j^A 
\sqrt{\xi - \sqrt{\xi^2 -1}}}
{2\xi} \simeq 
\frac{X_i ^A}{\sqrt{2 \xi}} - \frac{X_j^A}{(2\xi)^{3/2}}
,
\\
C_+ &= \frac{X_j ^A \sqrt{\xi + \sqrt{\xi^2 -1}} - X_i^A 
\sqrt{\xi - \sqrt{\xi^2 -1}}}
{2\xi} \simeq
\frac{X_j ^A}{\sqrt{2 \xi}} - \frac{X_i^A}{(2\xi)^{3/2}}
,
\end{align}
where we also expanded $C_{\pm}$ for $\xi \gg 1$, which is relevant when we put one of the points near the boundary. Now we will use this to calculate the minimal distance between spacelike geodesics $X_{14}(\lambda),X_{23}(\lambda ')$, each anchored on one boundary (points 1,2) and one bulk point (points 4,3). It's important to note that while for the case of boundary-boundary spacelike geodesics we can just keep the leading terms in the expansion of $C_{\pm}$, for the case of bulk-boundary geodesics we have to also keep second terms in our calculation. The reason for that will become clear below.

\subsection{Minimal distance between bulk-boundary geodesics}
Following previous subsection we write bulk-boundary geodesics as
\begin{align}
\label{eq: X14X23series}
X_{14} ^A (\lambda) &= e^{-\lambda}
\left(
\frac{X_1 ^A}{\sqrt{2\xi_{14}}}
-
\frac{X_4 ^A}{(2\xi_{14})^{3/2}}
\right) + 
e^{\lambda}
\left(
\frac{X_4 ^A}{\sqrt{2\xi_{14}}}
-
\frac{X_1 ^A}{(2\xi_{14})^{3/2}}
\right) ,
\\
X_{23} ^A (\lambda ') &= e^{-\lambda'}
\left(
\frac{X_2 ^A}{\sqrt{2\xi_{23}}}
-
\frac{X_3 ^A}{(2\xi_{23})^{3/2}}
\right) + 
e^{\lambda'}
\left(
\frac{X_3 ^A}{\sqrt{2\xi_{23}}}
-
\frac{X_2 ^A}{(2\xi_{23})^{3/2}}
\right)
, \nonumber
\end{align}
with $\xi_{ij} = - X_i \cdot X_j$. For each of these geodesics, one point corresponds to the end of the boundary interval while the other corresponds to its respective intermediate point on the horizon. The distance between two points $X_{14}(\lambda)$ and $X_{23}(\lambda ')$ is given by
\begin{align}
\label{eq: distance14,23}
d (X_{14}(\lambda),X_{23}(\lambda ')) &= 
 \phantom{,} \textup{arccosh}(\xi) = 
 \log (\xi + \sqrt{\xi^2 -1}) , 
\\
\xi &= - X_{14}(\lambda) \cdot X_{23}(\lambda ') .
\end{align}
To find the minimal distance between these two geodesics we just need to optimize over $\lambda,\lambda'$. This reduces to optimization of $\xi$ in a general form 
\begin{equation}
\xi = e^{-\lambda - \lambda'} A_{11} +
e^{\lambda+\lambda'} A_{22} + 
e^{- \lambda+ \lambda'} A_{12} + 
e^{\lambda-\lambda'} A_{21} ,
\end{equation}
for which the minimal value is found to be
\begin{equation}
\label{eq: xi_min}
\xi_{min} = 2( \sqrt{A_{12} A_{21}} + \sqrt{A_{11} A_{22} }  ).
\end{equation}
For $X_{14}(\lambda),X_{23}(\lambda ')$ written as in (\ref{eq: X14X23series}), we have
\begin{align}
A_{11} &= 
\frac{\xi_{12}}{2\sqrt{\xi_{14}\xi_{23}}} 
+ \frac{\xi_{34}}{(4 \xi_{14} \xi_{23})^{3/2}}
- \frac{\xi_{13}}{\sqrt{ 2\xi_{14}} (2\xi_{23})^{3/2} }
- \frac{\xi_{24}}{\sqrt{ 2\xi_{23}} (2\xi_{14})^{3/2} }
,\\
A_{22} &= 
\frac{\xi_{34}}{2\sqrt{\xi_{14}\xi_{23}}} 
+ \frac{\xi_{12}}{(4 \xi_{14} \xi_{23})^{3/2}}
- \frac{\xi_{24}}{\sqrt{ 2\xi_{14}} (2\xi_{23})^{3/2} }
- \frac{\xi_{13}}{\sqrt{ 2\xi_{23}} (2\xi_{14})^{3/2} }
,\\
A_{12} &= 
\frac{\xi_{13}}{2\sqrt{\xi_{14}\xi_{23}}} 
+ \frac{\xi_{24}}{(4 \xi_{14} \xi_{23})^{3/2}}
- \frac{\xi_{12}}{\sqrt{ 2\xi_{14}} (2\xi_{23})^{3/2} }
- \frac{\xi_{34}}{\sqrt{ 2\xi_{23}} (2\xi_{14})^{3/2} }
,\\
A_{21} &= 
\frac{\xi_{24}}{2\sqrt{\xi_{14}\xi_{23}}} 
+ \frac{\xi_{13}}{(4 \xi_{14} \xi_{23})^{3/2}}
- \frac{\xi_{34}}{\sqrt{ 2\xi_{14}} (2\xi_{23})^{3/2} }
- \frac{\xi_{12}}{\sqrt{ 2\xi_{23}} (2\xi_{14})^{3/2} }
.
\end{align}
Now because points (1,2) are at the boundary, we can drop the terms which lead to subleading $O(r^{-1})$ contributions to $\xi$. The relevant terms are 
\begin{align}
A_{11} &= 
\frac{\xi_{12}}{2\sqrt{\xi_{14}\xi_{23}}} 
\sim O(r)
,\\
A_{22} &= 
\frac{\xi_{34}}{2\sqrt{\xi_{14}\xi_{23}}} 
+ \frac{\xi_{12}}{(4 \xi_{14} \xi_{23})^{3/2}}
- \frac{\xi_{24}}{\sqrt{ 2\xi_{14}} (2\xi_{23})^{3/2} }
- \frac{\xi_{13}}{\sqrt{ 2\xi_{23}} (2\xi_{14})^{3/2} }
\sim O(r^{-1})
,\\
A_{12} &= 
\frac{\xi_{13}}{2\sqrt{\xi_{14}\xi_{23}}} 
- \frac{\xi_{12}}{\sqrt{ 2\xi_{14}} (2\xi_{23})^{3/2} }
\sim O(r^{0})
,\\
A_{21} &= 
\frac{\xi_{24}}{2\sqrt{\xi_{14}\xi_{23}}} 
- \frac{\xi_{12}}{\sqrt{ 2\xi_{23}} (2\xi_{14})^{3/2} }
\sim O(r^{0})
.
\end{align}
Note that even though $A_{22}$ is of the order $O(r^{-1})$, in combination with $A_{11}$ it gives the leading contribution to $\xi_{min}$. Inserting the endpoints of the geodesics to $\xi$'s leads to entanglement wedge cross-section (putting back $R$)
\begin{align}
E_W &= \frac{d (X_{14},X_{23})}{4G_N} \\
&= \frac{1}{4G_N} \textup{arccosh} \left(
\frac{2 \cosh R(x_A - x_B) + h(x_A) + h(x_B)}
{2 \sqrt{(h(x_A)+1)(h(x_B)+1)}} \right) \\
&= \frac{c}{6} \textup{arccosh} \left(
\frac{2 \cosh \frac{2\pi}{\beta}(x_A - x_B) + h(x_A) + h(x_B)}
{2 \sqrt{(h(x_A)+1)(h(x_B)+1)}}
\right) ,
\end{align}
where in the last line we have used the holographic dictionary~\cite{brown1986}
\begin{equation}
c = \frac{3}{2G_N}, \phantom{aaa} R = \frac{2\pi}{\beta} .
\end{equation}
Analyzing the form of $h(x)$
\begin{equation}
h(x_A) = \frac{4\pi G_N h_\psi}{\sin \frac{2\pi}{\beta}\tau} e^{\frac{2\pi}{\beta}(t+x-x_A)} = 
\frac{12 \pi h_\psi i}{c(e^{2\pi i\tau/\beta}
-e^{-2\pi i\tau/\beta})} e^{\frac{2\pi}{\beta}(t+x-x_A)},
\end{equation}
we see that after setting $\epsilon = -\tau$, this result matches precisely with the one derived on the CFT side. This is the main result of our work. Below we will briefly analyze the functional dependence of $E_W (t)$ and find an interesting plateau behaviour for times $t>x_B - x$.

\subsection{Discussion of the result}
\label{sec: discussion of localized}
\begin{figure}
    \centering
    \includegraphics[width=7.4cm]{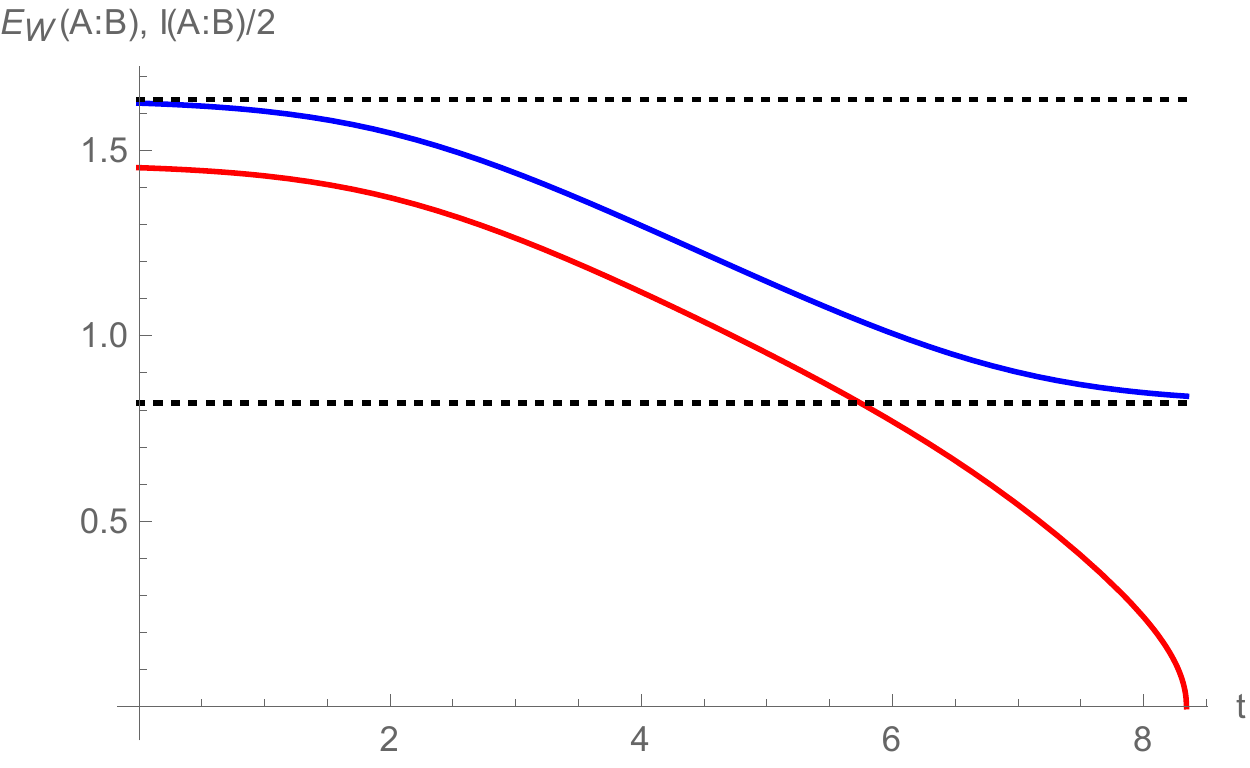}
    \includegraphics[width=7.4cm]{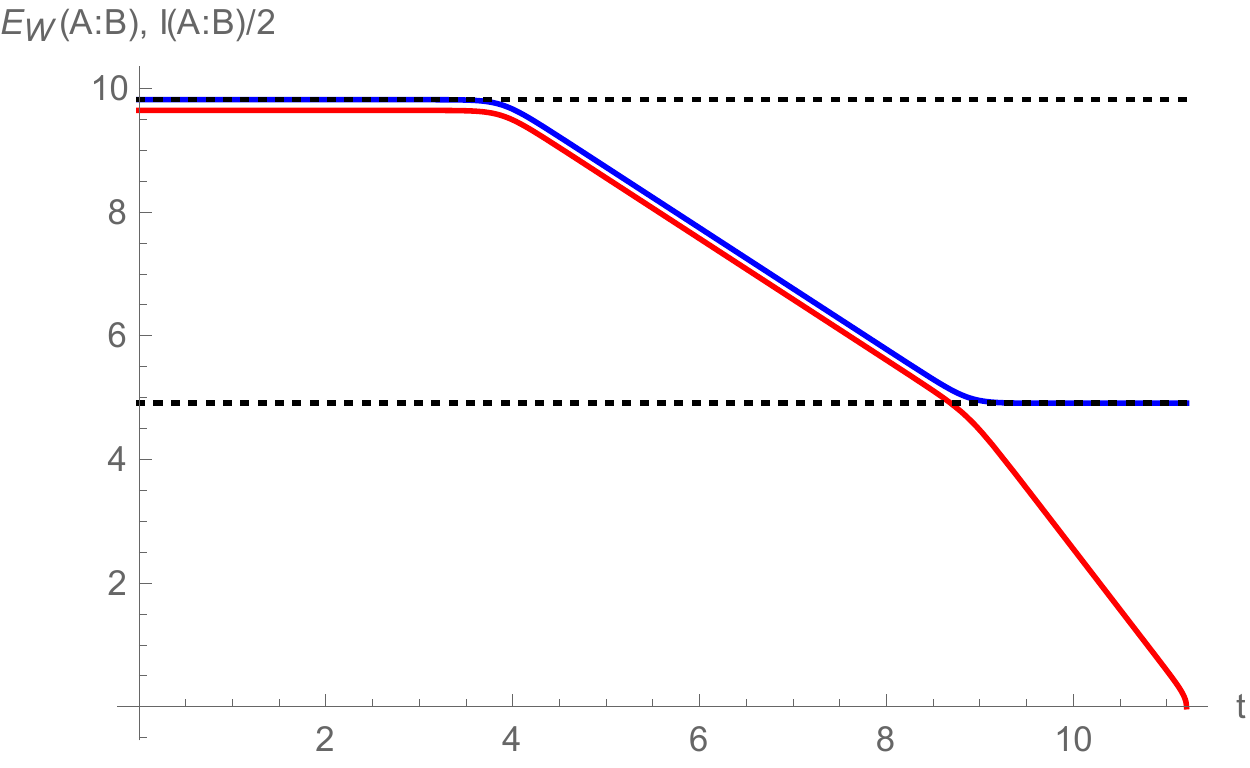}
    \caption{Half of the mutual information (red) and the entanglement wedge cross-section (blue) for localized shock wave for two different temperatures, with $t$ denoting the time in the past of the shock insertion. After the red line reaches zero, the entanglement wedge becomes disconnected and the blue line jumps to zero. In the above plots we took $x=1$, $x_A =5$, $x_B =10$. The dotted lines correspond to values $s_{eq}L$ and $s_{eq}L/2$ respectively. For high enough temperatures (right plot) an interesting plateau develops in the late time region, which should be properly captured by our setup. For low temperatures (left plot) mutual information disappears before we enter our region of validity, we therefore expect that there might be some corrections to the above behaviour.}
    \label{fig:ewcs_local}
\end{figure}

The result we arrived at
\begin{align}
E_W &= \frac{c}{6} \textup{arccosh} \left(
\frac{2 \cosh \frac{2\pi}{\beta}(x_A - x_B) + h(x_A) + h(x_B)}
{2 \sqrt{(h(x_A)+1)(h(x_B)+1)}}
\right) , \\
h(x_{A/B}) &=  
\frac{6 \pi h_\psi }{c \sin \frac{2\pi \tau}{\beta}} e^{\frac{2\pi}{\beta}(t+x-x_{A/B})},
\end{align}
should be compared with mutual information, to find when entanglement wedge becomes disconnected and EWCS goes to zero. Using the previously found HRT surfaces, we easily find
\begin{align}
I_{AB} &= \frac{1}{4G} \left( 2 \log 2r_\infty ^2 (\cosh \frac{2\pi}{\beta}(x_A - x_B)-1) - 2 \log 4 r_\infty ^2 (1+h(x_A))(1+h(x_B)) \right) \\
&= \frac{1}{2G} \log \frac{\cosh \frac{2\pi}{\beta}
(x_A - x_B)-1}{2 \sqrt{(h(x_A)+1)(h(x_B)+1)}} .
\end{align}
The comparison is plotted in Figure \ref{fig:ewcs_local} for different temperatures.
We see that in both cases the inequality~\cite{Umemoto_2018,Nguyen_2018}
\begin{equation}
E_W \geq \frac{1}{2} I_{AB}
\end{equation}
is properly satisfied. Since in both CFT and gravity calculations we assumed $t>x_B -x $, there might be some corrections to our result for times $t<x_B -x$. We expect however that for $t>x_B -x $ our result correctly captures the true behaviour. Certainly, the most interesting aspect of this result is the plateau which develops for high enough temperatures. It starts to develop after $h(x_B) = 1$, and settles at the value equal exactly half of the non-perturbed result
\begin{equation}
E_W ^{plat} = \frac{\pi}{4G \beta} (x_B - x_A ) = \frac{1}{2} s_{eq} L ,
\end{equation}
with $L=x_B -x_A$. We can in fact find the condition necessary for the plateau to develop. Denoting $t_p$ as the time at which $h(x_B)=1$, and $t_s$ the time for which $I_{AB} (t_s)=0$, a natural condition would be
\begin{equation}
t_p < t_s ,
\end{equation}
i.e. we want $h(x_B)$ to become significant before the entanglement wedge is disconnected. From this condition, we obtain 
\begin{equation}
\frac{L}{2} + \frac{\beta}{2\pi} \log \left(1-2 e^{-\frac{2\pi}{\beta} \frac{L}{2}} - 2e^{-\frac{2\pi}{\beta} \frac{3L}{2}}-e^{-\frac{2\pi}{\beta} 2L} \right) 
> \frac{\beta}{2\pi} \log 4 ,
\end{equation}
which for high temperatures $\beta \ll 1$ reduces to
\begin{equation}
\frac{\pi L}{2 \beta} > \log 2 \phantom{aaa} \Leftrightarrow 
\phantom{aaa}
\frac{1}{2} s_{eq} L > \frac{c}{3} \log 2 .
\end{equation}
This suggests that the plateau will develop only for temperatures bigger than $T_{c} \propto 1/L$.

A plateau with the same value of reflected entropy was also observed in~\cite{Kudler_Flam_2020} for global homogeneous quenches. There the authors found that the high-temperature behaviour of reflected entropy can be given a nice intuitive explanation by employing a line-tension picture~\cite{Nahum_2017,jonay2018coarsegrained,Mezei_2018,von_Keyserlingk_2018,Zhou_2019,Kudler_Flam_2020_linetension_negativity,Wang_2019}. In our case, we can use a generalization of this picture for local operator quenches in the context of reflected entropy, introduced by the same authors in a more recent work~\cite{kudlerflam2020entanglement}. We find 
\begin{equation}
E_W = 
\begin{cases}
L \log q & t<x_A -x, \\
\left(L - \frac{1}{2}(t+x-x_A)\right) \log q & x_A -x<t <x_B -x , \\
\frac{1}{2} L \log q & x_B -x< t < \frac{3}{2}x_B - \frac{1}{2}x_A -x , \\
0 & t> \frac{3}{2}x_B - \frac{1}{2}x_A -x .
\end{cases}
\end{equation}
This result precisely agrees with the high-temperature behaviour of our result after setting the bond dimension~\cite{Cardy:1986ie}
\begin{equation}
q = e^{s_{eq}} = e^{\frac{\pi c}{3 \beta}} .
\end{equation}

\section{Entanglement wedge cross-section for spherically symmetric shocks}
\label{sec:EWCS spherical shock wave}
\begin{figure}
    \centering
    \includegraphics[width=10cm]{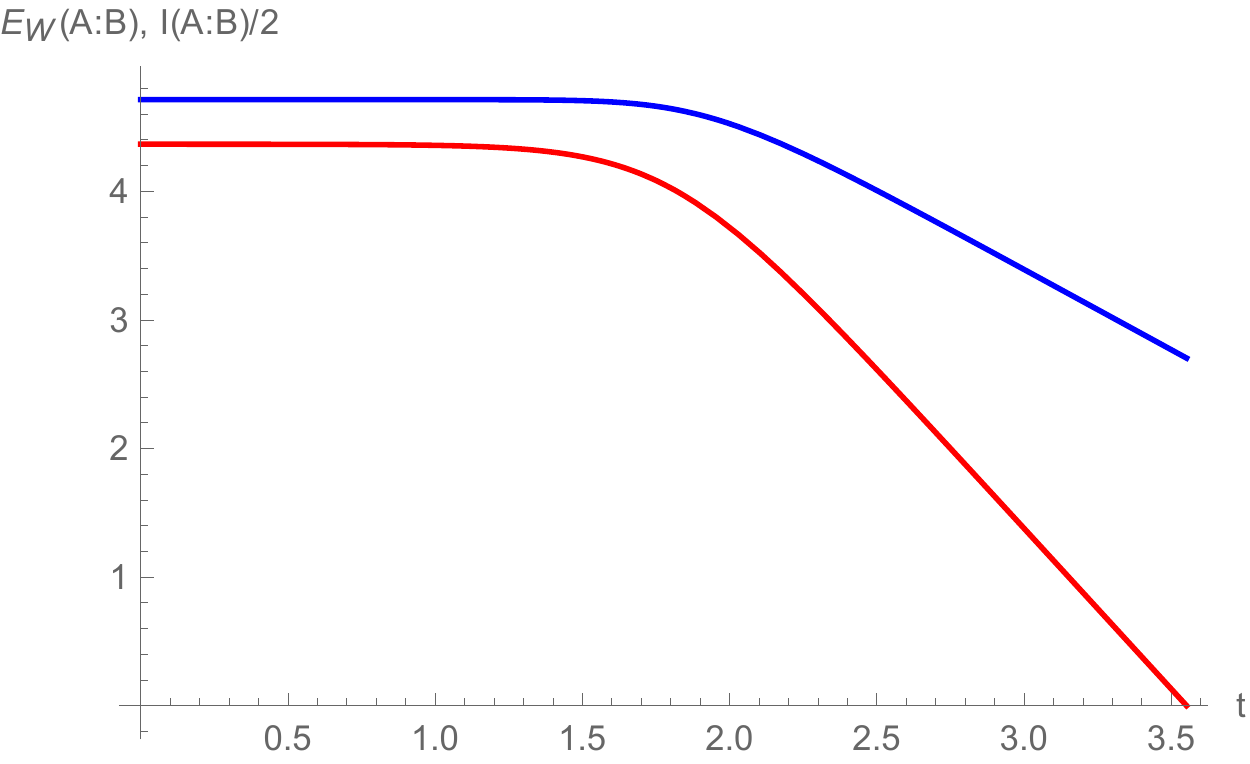}
    \caption{Half of the mutual information (red) and entanglement wedge cross-section (blue) for spherical shock waves, with $t$ denoting the time in the past of shock insertion. After the red line reaches zero, the entanglement wedge becomes disconnected and the blue line jumps to zero.}
    \label{fig:ewcs_spherical}
\end{figure}

Using the methods presented in Section \ref{sec:EWCS_localized shock wave}, we can also compute the entanglement wedge cross-section in the case of spherically symmetric null matter falling towards the AdS black hole~\cite{Shenker_2014}. We consider several particles with total energy $E$, thrown towards the AdS black hole with mass $M$ at time $t_w$ in the past. The metric in this case is\footnote{See \cite{Sfetsos_1995,Dray:1985yt,Hotta:1992qy} for more details on geometries with spherical shock waves.}
\begin{equation}
ds^2 = \frac{-4 du dv}{(1+uv)^2} + \frac{4 \alpha \delta(u)du^2}{(1+uv)^2} + R^2 \frac{(1-uv)^2 }{(1+uv)^2} d\phi^2 , \end{equation}
with
\begin{equation}
\alpha = \frac{E}{4M} e^{R t_w}. 
\end{equation}
This metric can also be thought of as two halves of AdS black hole with a shift in $v$ coordinate at $u=0$ 
\begin{equation}
v_L = v_R + \alpha .
\end{equation}
To find an entanglement wedge cross-section of two matching intervals of size $\phi$ at times $t_L=t_R=0$, we again start with finding the intermediate points on the horizon of HRT surfaces. The details of the computation are similar to the previous section, we find the minimal distance
\begin{equation}
d= \log \left(\frac{4r_\infty ^2}{R^2} \left( 1+ \frac{\alpha}{2} 
\right) \right) ,
\end{equation}
for HRT passing through 
\begin{equation}
v_R = -\frac{\alpha}{2}.
\end{equation}
Using now the formulas (\ref{eq: distance14,23}), (\ref{eq: xi_min}), for the radial geodesics between interval endpoints and the corresponding intermediate points on the horizon, we find 
\begin{equation}
E_W = \frac{1}{4G} \textup{arccosh} \left( 
1+ \left(\cosh R\phi -1\right)
\frac{4  \left(1 + \alpha \right)}{\left(\alpha + 2 \right)^2}
\right) .
\end{equation}
Note that setting $\alpha=0$ we reproduce the thermofield double result derived in~\cite{Nguyen_2018}. The comparison with half of the mutual information is plotted in Figure \ref{fig:ewcs_spherical}. In this case, there is no late time plateau region and entanglement wedge cross-section behaves similarly to mutual information before going to zero. 

We expect that this result should be also properly captured by the line-tension picture, however one would need to extend this prescription to capture the case of the spherical shock wave. We leave this for future work.

\section{Summary and discussion}
\label{sec:discussion}
In this work, we studied dynamical aspects of the entanglement wedge cross-section. On the CFT side, we used two conformal maps to derive reflected entropy in a thermofield double state perturbed by a sufficiently early heavy operator insertion. We showed that the result matches precisely with the gravity result derived from localized shock wave geometry~\cite{Roberts_2015,Roberts_2015_shocks}. For high temperatures, the result exhibits an interesting plateau behaviour in the time region which should be properly captured by our calculation. On the gravity side, we can see that the plateau develops because HRT surfaces can avoid the shock in the spatial direction. The value of the plateau is equal to exactly half of the EWCS in the unperturbed thermofield double state. We estimated that this plateau can develop only if its value is bigger than $\frac{c}{3} \log 2$. Employing a line-tension picture for local operator quenches~\cite{kudlerflam2020entanglement} we found that it precisely reproduces the high-temperature behaviour of our result.

We extended our calculation to the setup of spherically symmetric shock waves~\cite{Shenker_2014,Sfetsos_1995}. In this case, the shock is not localized and the HRT surfaces cannot avoid the shock by bending in the spatial direction. Here EWCS behaves similarly to mutual information and no plateau is observed. To our knowledge, the line-tension picture has not been yet extended to these situations, however, we expect that such a picture should properly capture the high-temperature behaviour of this setup.

A similar plateau region for reflected entropy was recently observed in~\cite{Kudler_Flam_2020} in the case of global homogeneous quenches. There the authors referred to it as "missing entanglement" $S_R - I$. This is because, as we've seen, the extended plateau period is not captured by mutual information. In our work, we have provided one situation in which the plateau is observed - localized shock wave, and one in which it is not - spherical shock wave. It would be interesting to understand what exactly is responsible for the plateau. Extending the line-tension picture to capture spherical shocks would probably be the next step in this direction.

\acknowledgments
I am grateful to Kotaro Tamaoka and Tadashi Takayanagi for comments on the draft and useful discussions. In particular, I would like to thank Pawel Caputa for careful reading of the draft, his continuous mentorship and support. I am also grateful to Yukawa Institute for Theoretical Physics for hospitality during the early stages of this work.

\appendix

\section{Embedding coordinates}
\label{app:conventions}
In our conventions AdS$_3$ space is defined in flat space $\mathbb{R}^{2,2}$
\begin{equation}
ds^2 = -dX_0 ^2 - dX_1 ^2 + dX_2 ^2 + dX_3 ^2 ,
\end{equation}
as the surface defined by  
\begin{equation}
- X_0 ^2 - X_1 ^2 + X_2 ^2 + X_3 ^2 = - 1 .
\end{equation}
The parametrization via Kruskal coordinates $(u,v)$ is given by
\begin{align}
X_0 &= \frac{v+u}{1+uv} 
,\\
X_1 &= \frac{1-uv}{1+uv} \cosh{R x}
,\\
X_2 &= \frac{v-u}{1+uv}
,\\
X_3 &= \frac{1-uv}{1+uv} \sinh{R x}
,
\end{align}
with $R$ denoting the horizon radius. Both sides of this geometry outside of the horizon can be parametrized by AdS-Schwarzschild coordinates $(r,t_{L/R})$ which are related to Kruskal coordinates as 
\begin{align}
\textup{Left side}:& \phantom{aa} u= \sqrt{\frac{r-R}{r+R}} e^{-R t_L},
\phantom{aa} v= -\sqrt{\frac{r-R}{r+R}} e^{R t_L}
,\\
\textup{Right side}:& \phantom{aa} u= -\sqrt{\frac{r-R}{r+R}} e^{-R t_R},
\phantom{aa} v= \sqrt{\frac{r-R}{r+R}} e^{R t_R}
.
\end{align}
In the above $t_L$ increases "downwards". The embedding coordinates allow us to easily calculate geodesic distances between points $X^i , X^f$ as
\begin{equation}
\cosh d = X_0 ^i X_0 ^f + X_1 ^i X_1 ^f - X_2 ^i X_2 ^f -X_3 ^i X_3 ^f .
\end{equation}

\bibliography{biblio.bib}

\end{document}